\documentclass{aa}
\usepackage{natbib}
\usepackage{graphicx}
\begin{document}
\title{{\it Research note}\\
Is there a phase constraint for solar dynamo models? }
\titlerunning{Is there a phase constraint for solar dynamo models?}
\author{M. Sch\"ussler}
\institute{Max-Planck-Institut f\"ur
Sonnensystemforschung,
Max-Planck-Str. 2, 37191 Katlenburg-Lindau, Germany 
}


\date{Received; accepted} 

\abstract{The spatio-temporal relationship between the sign of the
observed radial component of the magnetic field at the solar surface and
the sign of the toroidal field as inferred from Hale's polarity rules
for sunspots is usually interpreted as signifying the phase relation
between the poloidal and the toroidal magnetic field components involved
in the solar dynamo process. This has been taken as a constraint for
models of the solar dynamo. This note draws attention to the fact that
the observed phase relation is naturally and inevitably produced by the
emergence of tilted bipolar regions and flux transport through surface
flows, without any necessity of recourse to the dynamo
process. Consequently, there is no constraint on dynamo models resulting
from the observed phase relation.
}
\maketitle


The relationship between the sign of the observed longitude-averaged
radial magnetic field component in the photosphere, $B_r$, and the
sign of the azimuthal field, $B_\phi$, as inferred from the polarities
of the following and preceding parts of sunspot groups (according to
Hale's rules), is taken by many authors as an important
constraint for models of the solar dynamo \citep[e.g.,][]{Stix:1976,
Yoshimura:1976, Parker:1987b, Schmitt:1993,
Schlichenmaier:Stix:1995, Ruediger:Brandenburg:1995,
Bonanno:etal:2002, Ossendrijver:2003, Brandenburg:2005}.

\begin{figure}
  \centering 
  \resizebox{\hsize}{!}{\includegraphics{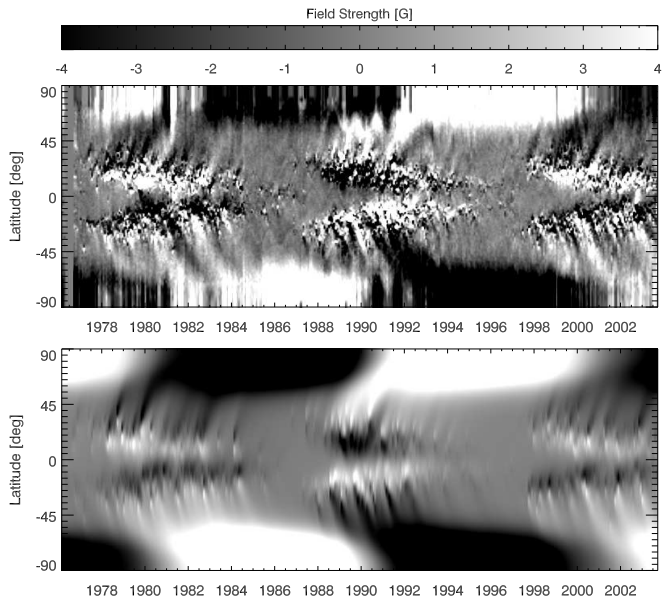}}
\caption{Comparison between observed and simulated time-latitude plots
(butterfly diagram) of the longitudinally averaged radial magnetic field
at the solar surface. {\it Upper panel:} Evolution of the observed
field, based upon NSO Kitt Peak synoptic maps (courtesy
D. Hathaway). {\it Lower panel:} Simulation for the same period of time
with the flux transport code of \citet{Baumann:etal:2004}. The emerging
active regions have been determined from the RGO and NOOA/USAF SOON
sunspot data. In both cases, the grey scale is saturated at 4 G to
better bring out the low-latitude fields.  The dominance of the
leading-polarity flux in low latitudes (corresponding to the phase
relation $B_r B_\phi < 0$) due to the tilt angle of the emerging active
regions is reproduced by the flux-transport simulation, which does not
involve any assumption about the working of the dynamo.  The phase
relation therefore cannot be taken as a constraint on dynamo models.}
\label{fig:butterfly}
\end{figure}

\citet{Stix:1976} considered Mount Wilson magnetograph data for the
period 1959--1973.  Taking $B_r$ positive in the outward radial
direction and $B_\phi$ positive in the direction of solar rotation, he
found the relation $B_r B_\phi < 0$ to hold in the sunspot zones below
35 degrees heliolatitude. In a time-latitude diagram, the average radial
field shows `butterfly wings' that closely match the corresponding
sunspot pattern. The `phase relation', $B_r B_\phi < 0$, then means that
the average radial field has the same polarity as the preceding parts of
the active regions emerging during the same half cycle \citep[][see also
the upper panel of Fig.~\ref{fig:butterfly}]{Stenflo:1972,
Howard:Labonte:1981, Schlichenmaier:Stix:1995}.

When taking the phase relation as a constraint for solar dynamo models,
the tacit assumption is made that the observed radial field at the
surface actually represents the poloidal field component resulting from
the (deep-seated) dynamo process. However, there is evidence that this
assumption is not necessarily valid. The observed evolution of sunspots
and active regions indicate that the corresponding magnetic flux is
dynamically disconnected from its subsurface roots within a few days
after emergence \citep{Fan:etal:1994, Schrijver:Title:1999,
Schuessler:2005, Schuessler:Rempel:2005}. In fact, the large-scale
evolution of the observable magnetic flux at the solar surface is
reproduced quite well by the so-called flux-transport models, which
simulate the passive advection of the radial magnetic field by the
near-surface flows of supergranulation (described as a turbulent
diffusion process), differential rotation, and meridional circulation
\citep[e.g.,][]{Wang:etal:1989, Schrijver:2001, Mackay:etal:2002,
Baumann:etal:2004}.  The flux input in such models is provided by the
emergence of bipolar magnetic regions, taken either directly from the
observational data or assumed at random locations, but keeping the basic
statistical properties of active regions (tilt angle according to Joy's
law, latitude drift of the activity belt, Hale's polarity rules). As an
illustration, Fig.~\ref{fig:butterfly} shows a comparison of the result
from the flux-transport model of \citet{Baumann:etal:2004} with the
actual time-latitude diagram of the (longitudinally averaged) observed
surface field. The input for the flux-transport simulation has been
derived from the RGO and NOOA/USAF SOON sunspot data (Baumann et
al. 2005, in preparation).

It turns out that all such flux-transport models reproduce the observed
phase relation, $B_r B_\phi < 0$, in low latitudes, i.e., they all show
that the longitude-averaged radial field predominantly has the same
polarity as the leading parts of the active regions emerging throughout
the same half cycle \citep[see also Fig.~3 of][]{Baumann:etal:2004}. No
assumption about the dynamo and, in particular, about the phase relation
between the poloidal and toroidal field components in the dynamo process
is required to obtain this result. In fact, the basic assumption
underlying these models is that the large-scale radial surface flux
results exclusively from the local emergence of active regions. The
dominance of the leading-polarity flux in low latitudes arises from the
tilt of the bipolar regions: the leading parts are nearer to the
equator, so that they dominate, on average, the low latitudes. This
effect is amplified by the latitude gradient of the poleward meridional
flow speed, which leads to a preferential poleward transport of the
(opposite-polarity) following parts of the bipolar regions. The same
result is found in the complimentary model of
\citet{Choudhuri:Dikpati:1999}, who consider longitude-averaged
quantities in the meridional $(r,\theta)$ plane: the poloidal field
resulting from tilted bipolar regions reproduces the observed
phase relationship with the toroidal field component, in the absence of
any assumptions concerning the dynamo process.

As a side remark, we note that in the class of Babcock-Leighton-type
advection-dominated dynamos \citep[e.g.,][]{Dikpati:Charbonneau:1999b}
the tilt of the bipolar magnetic regions provides the source of the
poloidal field for the {\it next\/} (half) cycle. In that sense, this
kind of dynamos automatically reproduces the observed phase relation,
but that does not exclude other dynamo models that do not rely on the
tilt of active regions as the source for the poloidal field. Note also
that Joy's law for the tilt angle of bipolar magnetic regions is
explained completely independent from any dynamo model by the action of
the Coriolis force on rising flux loops
\citep[e.g.,][]{Dsilva:Choudhuri:1993, Fan:etal:1994,
Caligari:etal:1995}.

Several authors have pointed out that the tilt angle of active
regions leads to a dominance of leading polarity flux in the sunspot
latitudes \citep[e.g.,][]{Stenflo:1972, Howard:Labonte:1981,
Wang:Sheeley:1991b, Choudhuri:Dikpati:1999}.  The message that this
provides a natural explanation for the phase relation independent of the
dynamo process apparently has not reached the majority of the dynamo
community. It is the purpose of this note to draw attention to this
result and thus prevent the exclusion of dynamo models because of
an inappropriate constraint.

\acknowledgement{Ingo Baumann kindly provided Fig.~\ref{fig:butterfly}.}

\bibliographystyle{aa}
\bibliography{3459.bbl}

\end{document}